\newif\ifarxiv
\arxivtrue
\ifarxiv
\pdfoutput=1
\documentclass[10pt]{article}
\newcommand{\name}[1]{#1\\}
\newcommand{\affil}[1]{\parbox{\textwidth}{\small#1}}
\long\def\tbl#1#2{#2\caption{#1}}
\else
\documentclass{interact}
\fi
\usepackage{amsmath}
\usepackage{amsthm}
\usepackage{amssymb}
\usepackage{bm}  
\usepackage{url}
\usepackage[numbers]{natbib}
\usepackage{enumerate}
\usepackage{graphicx}
\usepackage{blkarray}
\usepackage{color}
\usepackage{bbm}
\usepackage{dsfont}
\usepackage{subcaption}
\usepackage[toc,page]{appendix}
\usepackage{blkarray}
\newcommand{\glpsol}{\textsc{glpsol}}
\newcommand{\cplex}{\textsc{cplex}}
\newcommand{\gurobi}{\textsc{gurobi}}

\newcommand{\Pp}{{\textsc{P}}}
\newcommand{\NP}{{\textsc{NP}}}

\newcommand{\Ppoly}{{\textsc{P/poly}}}

\newcommand{\RR}{{\mathbb{R}}}

\newcommand{\EP}{\mathrm{EP}}

\DeclareMathOperator{\CH}{CH}

\newif{\ifshowchanges}
\ifarxiv
\showchangesfalse
\else
\showchangestrue
\fi
\ifshowchanges
\newcommand{\change}[1]{\textcolor{blue}{#1}}
\usepackage[showchanges]{sparklistings}
\else
\newcommand{\change}[1]{#1}
\usepackage{sparklistings}
\fi

\usepackage{textcomp}
\usepackage{amssymb}
\usepackage{fullpage}

\usepackage{tcolorbox}
\tcbuselibrary{listings}
\newtcblisting{sparksbox}[1]{listing only,listing remove caption=false,colback=black!3,listing options={style=change,frame=none,#1}}
\DeclareMathOperator{\steps}{steps}
\newcommand{\Lower}{\codevar{lower}}
\newcommand{\Upper}{\codevar{upper}}
\newcommand{\Body}{\codevar{body}}
\newcommand{\BoolExpr}{\codevar{bool\_expr}}
\newtcolorbox{spasm}[1][]{colback=black!3,colframe=black,colbacktitle=black!10,coltitle=black,sidebyside,lefthand width=2in,title=#1}
\newcommand{\Sparks}{\texttt{Sparks}}
\newcommand{\Sparktope}{\textsc{sparktope}}
\newcommand{\Asm}{\texttt{Asm}}

\newcommand{\codevar}[1]{\textsf{#1}}

\usepackage[bordercolor=white]{todonotes}

\title{\textsf{Sparktope}: linear programs from algorithms}

\author{\name{David Avis\textsuperscript{a} and David Bremner\textsuperscript{b}
    \thanks{\textsc{Contact} D.\ Bremner. Email: bremner@unb.ca}}
\affil{\textsuperscript{a}School of Informatics, Kyoto University, Kyoto,
       Japan and 
       School of Computer Science,
       McGill University, Montr{\'e}al, Qu{\'e}bec, Canada;
       \textsuperscript{b}
       Faculty of Computer Science,
       University of New Brunswick, Fredericton, New Brunswick, Canada}}

\theoremstyle{plain}
\newtheorem{theorem}{Theorem}

\newtheorem{proposition}{Proposition}

\theoremstyle{definition}
\newtheorem{definition}{Definition}

\theoremstyle{remark}

\BAnewcolumntype{s}{>{$\small$}c}

\newlength{\normalparindent}
\newlength{\normalparskip}
\begin{document}

\maketitle


\begin{abstract}
In a recent paper Avis, Bremner, Tiwary and Watanabe gave a method for constructing
linear programs (LPs) based on algorithms written in a simple programming language
called \Sparks. 
If an algorithm produces the solution $x$ to a problem in polynomial time and space
then the LP constructed is also of polynomial size and its optimum solution contains $x$
as well as a complete execution trace of the algorithm.
Their method led us to the construction of a compiler called \Sparktope~which we
describe in this paper. 
This compiler allows one to generate polynomial sized LPs for problems in $\Pp$ that have
exponential extension complexity, such as matching problems in non-bipartite graphs.

In this paper we describe \Sparktope, the language \Sparks,
and the assembler instructions and LP constraints it produces.
This is followed by two concrete examples, the makespan problem and the problem of
testing if a matching in a graph is maximum, both of which are known to have exponential
extension complexity. Computational results are given.
In discussing these examples we make use of visualization techniques
included in \Sparktope~that may be of independent interest. 
The extremely large linear programs produced by the compiler appear to be quite challenging
to solve using currently available software. Since the optimum LP solutions can be computed
independently they may be useful as benchmarks.
Further enhancements of the compiler and its application are also
discussed.

\ifarxiv
\fi
\end{abstract}

\ifarxiv
\relax
\else
\fi

\section{Introduction}
\label{intro}
Linear programming is one of the big success stories of optimization and is routinely used to efficiently 
solve
extremely large problems. 
\change{
In 1982 Valiant~\cite{Valiant82} showed that any problem in $\Pp$ can be solved by
a polynomial sized linear program (LP). His proof technique, via families of circuits, was {\em uniform} in the sense that all
instances of the problem with given size $n$ can be solved by a single LP.
However, for a given computational problem, the construction of such families of circuits appears quite difficult. 
The question of how to systematically construct
these LPs lead to the recent work of Avis et al.~\cite{ABTW19} which we continue in this paper.
The main contribution of \cite{ABTW19} was to give a direct method to produce
polynomial
size LPs
from polynomial time algorithms.
Specifically they constructed LPs directly from
algorithms expressed in a simple language called \Sparks. The resulting LPs are uniform in the sense
described above. The main purpose of the present paper is to describe the implementation of a
compiler that takes \Sparks~code as input and produces an LP that performs the same function.
We give two concrete examples of its use: the {\em makespan problem} and the {\em maximum matching problem}.
}

\change{
The matching problem and its relationship to the
field of {\em extension complexity} was one of the main motivations of our work.
}
A matching $M$ in an undirected graph $G=(V,E)$ with $n$ vertices is a set 
of vertex disjoint edges from $E$.
The maximum matching problem is to find a matching in $G$ with the largest number of edges.
A related decision problem is to decide if a given matching $M$ in $G$ has maximum size.
Both of these problems can be solved in polynomial time by running Edmonds' algorithm \cite{Edmonds1965a}.
As well as this combinatorial algorithm, Edmonds also introduced a related polytope
 \cite{Edmonds1965b}
which is known as the {\em Edmonds' polytope} $\EP_n$ and whose variables
correspond the the $n(n-1)/2$ edges of the complete graph $K_n$. 
Matching problems can be solved by a linear program with constraint set $\EP_n$,  where the 
coefficients of the objective
function are one for the edge set $E$ and zero otherwise.
Unlike his algorithm's polynomial running time,
$\EP_n$
has size exponential in $n$. This raised the question of whether $\EP_n$ could
be the projection of a higher dimensional polytope that does have polynomial size.
Such a polytope is called an {\em extended formulation} and
is the central concept in {\em extension complexity} (see, e.g., Fiorini et al.~\cite{FMPTW15}).
However, in a celebrated result, Rothvoss \cite{Rothvoss17} proved 
$\EP_n$ admits no polynomial size extended formulation.
Rothvoss's result has sometimes been misinterpreted to mean that no polynomial sized
LP exists for the matching problem. 
\change{
In Section \ref{matching} of the paper we give details of the construction of a uniform family
of LPs for the maximum matching problem that have polynomial size.
}

\change{
The \Sparks~language
}
was modeled on Sahni's proof of Cook's
theorem given in \cite{HS78}. Since \Sparks~is strong enough to
implement Edmonds' algorithm in polynomial time, it can produce the required polynomial sized
LPs for the matching problem. In this paper we describe the implementation those ideas in a compiler 
we developed called \Sparktope. We then show how \Sparktope~can
be used to produce polynomial sized LPs for two problems with exponential extension 
complexity: makespan and maximum matching.
We should emphasize here that \Sparktope~will convert any algorithm written in \Sparks~
into a linear program. However this LP will only have polynomial size if the
algorithm terminates in polynomial time.
\change{
Nevertheless our approach gives new results even for NP-hard problems.
For example, converting the Held-Karp travelling salesman algorithm \cite{HK62} into an
LP via \Sparks~gives an asymptotically smaller LP than that formed from 
the convex hull of the Hamiltonian circuits. This is discussed further in
the concluding remarks.
}

The paper is organized as follows. In Section \ref{sect:poly} we summarize
the results in \cite{ABTW19} and their main theorem.
Sections \ref{stepcounter} and \ref{sect:general}
describe details of the \Sparks~compiler and how it
produces linear programs.
This is followed by Sections \ref{makespan} and \ref{matching} which describe 
the application of \Sparktope~to produce linear programs
for the makespan and maximum matching problems. 
We give some concluding remarks in Section \ref{sect:conclusions}.
In the appendices we 
give a complete \Sparks~code for the matching problem and a sample input. 

\section{Linear programs and weak extended formulations}
\label{sect:poly}
In this section we review the results in \cite{ABTW19} that are relevant to
the present paper. The proofs for results stated here can be found
in that paper. 
For simplicity we initially restrict the discussion to decision
problems however the results apply to optimization problems also.
Let $X$ denote a decision problem defined on binary input vectors $x=(x_1,...,x_q)$,
and an additional bit $w_x$, where $w_x=1$ if $x$ results in a ``yes'' answer and $w_x=0$
if $x$ results in a ``no'' answer. We define the polytope $P$ as:
\begin{equation}
\label{CH2}
P =\CH\{(x,w_x): x \in \{0,1\}^q\}
\end{equation}

For a given binary input vector $\bar{x}$ we define the vector $c=(c_{j})$ by:
\begin{equation}
\label{c2}
c_{j} = \begin{cases}
~~1 & \text{if}~\bar{x}_j=1\\
-1 & \text{if}~\bar{x}_j=0
\end{cases}
~~~~~~~~~~1 \leq i < j \leq n
\end{equation}
and let $d$ be a constant such that $0 < d \leqslant 1/2$.
We construct an LP:
\begin{eqnarray}
\label{LP3}
z^*  = &\max&~ c^Tx  + d w\\
&s.t.&~  (x,w) \in P \nonumber
\end{eqnarray}
\begin{proposition}[\cite{ABTW19}]
\label{optlemma2}
For any  $\bar{x} \in \{0,1\}^q$
let $m=\mathds{1}^T\bar{x}$.
The optimum solution to (\ref{LP3}) is unique, $z^* = m +d$ if $\bar{x}$ has a ``yes'' answer
and $z^* = m$ otherwise.
\end{proposition}
We will be interested in problems where the constraint set describing $P$ has
an exponential number of constraints implying that the LP (\ref{LP3}) has
exponential size.
It may still be the case that $P$ is
be the projection of a higher dimensional polytope $Q$ that does have polynomial size,
which would give a polynomial size LP.
Two examples of this are the spanning tree polytope (see Martin~\cite{Martin91}) and the cut polytope for
graphs with no $K_5$ minor (see, e.g., Deza and Laurent~\cite{cutbook}, Section 27.3).
Such a polytope $Q$ is called an {\em extended formulation} and
is the central concept in {\em extension complexity} (see, e.g., Fiorini et al.~\cite{FMPTW15}).
The condition that $Q$ projects onto $P$ is a strong one: there are problems in $\Pp$
that have exponential extension complexity.
In a celebrated result Rothvoss \cite{Rothvoss17} proved that the maximum matching problem
in graphs was such a problem. We describe below how a weaker notion of extension complexity
leads to polynomial size LPs for problems in $\Pp$.
 
In what follows the $k$-cube refers to the $k$ dimensional hypercube whose vertices are all the
binary vectors of length $k$.
\begin{definition}[\cite{ABTW19}]
Let $Q$ be a polytope which is a subset of the $(q+t)$-cube
with variables labelled $x_1, ...,x_q$, $y_1,...,y_t$.
We say that $Q$ has the {\em x-0/1 property} if each of the $2^q$
ways of assigning 0/1 to the variables $x_1, ...,x_q$ uniquely extends to
a vertex $(x,y)$ of $Q$ and, furthermore, $y$ is 0/1 valued. 
$Q$ may have additional fractional vertices.
\end{definition}
In polyhedral terms,
for every binary vector $b \in \RR^q$, the intersection of $Q$ with the
hyperplanes $x_j = b_j$
is a 0/1 vertex of $Q$.
We will show that we can solve a decision problem $X$ 
by replacing $P$ in
(\ref{LP3}) by a polytope $Q$ based on an algorithm for solving $X$,
while maintaining the same objective 
functions. If this algorithm runs in polynomial time then $Q$ has
polynomial size.
We call $Q$ a {\em weak extended formulation}
as it does not
necessarily project onto $P$.
\begin{definition}[\cite{ABTW19}]
\label{def:wef}
A polytope 
\[
Q=\{(x,w,s) : x \in [0,1]^q, w \in [0,1],
s \in [0,1]^r, Ax + bw + Cs \leqslant h\} 
\]
is a {\em weak extended formulation (WEF)} of $P$ if the following hold:
\begin{itemize}
\item
$Q$ has the $x$-0/1 property. 
\item
For any vector $\bar{x} \in \{0,1\}^q$
let $m=\mathds{1}^T\bar{x}$, let $c$ be defined by (\ref{c2})
and let $0 < d \leqslant 1/2$.
If $\bar{x}$ has a ``yes'' answer \change{(i.e.\ $(\bar{x},1) \in P$)}
the optimum solution of the LP
\begin{equation}
\label{optlp}
z^*=\max~\{c^Tx + dw: (x,w,s) \in Q \}
\end{equation}
is unique and takes the value $z^* = m +d$.
\change{Otherwise the optimal solution may not be unique but $z^* < m+d$.}
\end{itemize}
\end{definition}
\noindent
The first condition states that any vertex of $Q$ that has 0/1 values for the
$x$ variables has 0/1 values for the other variables as well. The second
condition connects $Q$ to $P$. For a 0/1 valued vertex $(x,w,s)$ of $Q$
we have $w=1$ if $x$ encodes a ``yes'' answer since \change{$z^*=m+d=\mathds{1}^T\bar{x}$.} If $x$
encodes a ``no'' answer then $z^* < m+d$ and we must have $w=0$.
The purpose of the coefficient $d$ is so that we can distinguish the two
answers by simply observing the value of $z^*$.

In general $Q$ will have fractional vertices and that is why
the condition for ``no'' answers differs from that given in
Proposition \ref{optlemma2}.
However, for small enough $d$ we can ensure that the LP optimum solution is
unique in both cases and corresponds to that given in
Proposition \ref{optlemma2}.
\begin{proposition}[\cite{ABTW19}]
\label{aboutd}
Let $Q$ be a WEF of $P$. There is a positive constant $d_0$, 
whose size is polynomial in the size of $Q$,
such that for all $d$, $0<d < d_0$, the optimal solution of the LP 
defined in (\ref{optlp})
is unique, $z^* = m +d$ if $\bar{x}$ has a ``yes'' answer
and $z^* = m$ if $\bar{x}$ has a ``no'' answer. 
\end{proposition}

If we are able to observe the {\em value} of $w$ in the optimum solution of (\ref{optlp}) then
we may in fact set $d=0$. In this case 
$z^* = m$ for both answers and it follows from the 0/1 property that the optimum solution is unique
and 0/1 valued. Since $Q$ is a WEF of $P$
the value of $w$ in the optimum solution gives the correct answer to the decision problem.
This is the preferred method in practice as it reduces the problem of 
floating point round off errors
which may be caused by small values of $d$. 

Combining the above results with
the circuit complexity model the following theorem was obtained.
$\Ppoly$ is the class of decision problems that can be solved by polynomial size
circuits.

\begin{theorem}[\cite{ABTW19}]
\label{main_thm}
Every decision problem $X$ in $\Ppoly$ admits a weak extended formulation
$Q$ of polynomial
size.
\end{theorem}

Since constructing circuits is a cumbersome way to express algorithms the authors
obtained the same result by working with algorithms expressed in
pseudocode. This has the additional advantage that they could also obtain similar results
for optimization problems directly, i.e. without having to resort to binary search. 
They chose to use the language \Sparks~
which is described next.

\section{\Sparks~and its assembly language \Asm{}}
\label{stepcounter}

To convert an algorithm into an LP there is a trade-off between the ease of writing the
algorithm in a reasonably high level language and the ease of converting 
a program in this language
into a set of LP constraints. For this reason we follow the usual
practice of introducing a programming language and converting it to an
intermediate language (so-called ``assembly code'') before finally
converting the intermediate language into a set of LP constraints. We
detail the first two steps in this section and the third step in the
next section.
In order to get a single LP to handle all inputs of a given size $n$ it will be necessary
to get a bound on the number of steps required to complete any input of this size.
Since our step clock is based on assembly instructions executed it is necessary to
detail these for each high level instruction. 

The language \Sparks~was introduced in Horowitz and Sahni \cite{HS78} where it was used for
a proof of Cook's theorem that was based on pseudocode rather than Turing machines.
We have implemented those features of \Sparks~that are
necessary for expressing combinatorial algorithms such as
Edmonds' unweighted matching algorithm.
Additional features would be needed to handle more sophisticated problems, such
as the weighted matching problem. For further details, the reader is referred to
Section 11.2 of \cite{HS78}.

We refer to our intermediate language \Asm{} as an \emph{assembly
  code} since it implements a register based virtual instruction
set. Unlike a conventional compiler for a language like C or FORTRAN,
most of the translation effort is actually generating the final output
(in our case linear inequalities) from the \Asm{} code. Readers
familiar with virtual machine based languages like Java or
Python may find it helpful to think of the generated linear
constraints as implementing a virtual machine that executes the
\Asm{} instructions.

We distinguish between \emph{compile time}, when the system of
inequalities corresponding to a particular \Sparks{} program is
generated, and \emph{run time}, when the inputs to the program are
defined, and the resulting LP solved.

A \Sparks~program consists of a sequence of statements, where each
statement is either a variable declaration, an assignment, or a
block structured control statement.

\begin{itemize}
\item {\em Scalar variables} are binary valued or $W$-bit {\em
    integers}, for some $W$ fixed at compile time.

\item Arrays of binary values are allowed and may be one or two
  dimensional.  Dimension information is specified at the beginning of
  the program. One dimensional arrays of integers are equivalent to
  two dimensional binary arrays with $W$ columns.

\item \change{For an input size of $n$, we let $p(n)$ denote the maximum
  number of steps required for the program to complete
  and} $q(n)$ denote the maximum number of bits required to
  represent all variables. Sahni argues that
  $q(n)=O(p(n))$, however typically $q(n)$ is significantly smaller.

\item
  Certain variables are designated as \texttt{input} and used to
  provide input to the program at run time. All other variables are initially zero.

\item An \emph{assignment} has scalar variable or array reference on
  the left hand side, and an \emph{expression} on the right hand
  side. A \emph{simple expression} has a single operator (or is just a
  variable).  \Sparks{} supports a limited set of \emph{compound
    expressions}, currently only permitting joining two simple
  expressions with a binary operator.

\item \Sparks{} supports block structured \texttt{if},  \texttt{while}, and \texttt{for}
  statements.
\item
  The program terminates if it reaches a \texttt{return} statement, which sets a binary \texttt{output}
  variable as a side effect.
\end{itemize}

The remainder of this section gives
details of the above \Sparks{} statements along with the assembler code they generate. It is rather technical and may be skipped on first reading and referred to as necessary
to understand the examples given in Sections \ref{makespan} and \ref{matching}.

As noted above, the number of assembler code instructions is necessary to obtain
bounds on the size of the LP constraint set generated.
These bounds are given for each code sample below.
In these samples, $x,y,z$ are assumed declared boolean,
$i,j,k$ are assumed declared integer, $A$ is a boolean array, and $B$
is an integer array.
\subsection{Simple assignments}
\label{sec:simple}
\change{Some  \Sparks{} statements translate one-to-one to \Asm{} statements. We group them here according to argument type.}

\begin{spasm}[Boolean operations]
\begin{sparks}
z <- x
z <- !x
z <- x and y
z <- x or y
z <- x xor y
\end{sparks}
  \tcblower
\begin{asm}
. set z copy x
. set z not x
. set z and x y
. set z or x y
. set z xor x y
\end{asm}
\end{spasm}

\change{
In actual \Sparks{} input $\leftarrow$ is typed `\lstinline|<-|'. The
Boolean equality test is written `\lstinline|=|', rather than
`\lstinline|==|'. The `\lstinline|.|' in the first column of the
\Asm{} is a placeholder for a statement label.
}
\begin{spasm}[Integer operations]
\begin{sparks}
z <- x eq y
z <- i = j
z <- i < j
i <- j
i <- j + k
i <- inc(j)
i <- dec(j)
i++
\end{sparks}
  \tcblower
\begin{asm}
. set z eq x y
. set z eqw i j
. set z ltw i j
. set i copyw j
. set i addw j k
. set i incw j
. set i decw j
. set i incw i
\end{asm}
\end{spasm}
\begin{spasm}[Array operations]
\begin{sparks}
A[*] <- 0
B[*,*] <- 0
A[i] <- x
B[[i]] <- j
\end{sparks}
  \tcblower
\begin{asm}
. array_init A 0
. matrix_init B 0
. array_set A i x
. row_set B i j
\end{asm}
\end{spasm}
\change{The notation \lstinline[language=sparks]|B[[i]]| (entered as \lstinline[language={}]|B[[i]]|) indicates a
row of a Boolean matrix should be interpreted as an integer.}
\subsubsection{Negated operators}
Currently there is only one negated operator supported. It translates
to two \Asm{} statements. 
\begin{spasm}
\begin{sparks}
z <- i != j
\end{sparks}
  \tcblower
\begin{asm}
. set |_tmp| eqw i j
. set z not |_tmp|
\end{asm}
\end{spasm}

\subsection{Array reads}
\change{Array reads translate to one extra \Asm{} statement (and one extra
step) per array reference compared to the statements in
Section~\ref{sec:simple}.}
\begin{equation*}
  \steps(\codevar{array using expr}) = \steps(\codevar{basic expr}) + \#(\codevar{array refs}).
\end{equation*}

\begin{spasm}
\begin{sparks}
x <- A[i]
\end{sparks}
  \tcblower
\begin{asm}
. set  |_tmp| array_ref A i
. set x copy |_tmp|
\end{asm}
\end{spasm}

\begin{spasm}
\begin{sparks}
x <- A[i] and A[j]
\end{sparks}
  \tcblower
\begin{asm}
. set  |_tmp1| array_ref A i
. set  |_tmp2| array_ref A j
. set x and |_tmp1 _tmp2|
\end{asm}
\end{spasm}

\begin{spasm}
\begin{sparks}
j <- B[[i]]
\end{sparks}
  \tcblower
\begin{asm}
. set  |_tmp| row_ref B i
. set j copyw |_tmp|
\end{asm}
\end{spasm}

\begin{spasm}
\begin{sparks}
j <- B[[i]] + B[[k]]
\end{sparks}
  \tcblower
\begin{asm}
. set  |_tmp1| row_ref B i
. set  |_tmp2| row_ref B k
. set j addw |_tmp1 _tmp2|
\end{asm}
\end{spasm}

\subsection{Compound assignments}

For convenience \Sparks{} supports a single level of compound
expressions as right-hand-sides. In particular any right hand side
from Subsection~\ref{sec:simple} can be joined by an operator to any
other with matching type (i.e. \texttt{bool} or \texttt{int}). \change{The
steps needed for the resulting \Asm{} code can be calculated as
follows:}
\begin{equation*}
  \steps(\codevar{compound})=\steps(\codevar{rhs}) + \steps(\codevar{lhs}) + 1\,.
\end{equation*}

\begin{spasm}
\begin{sparks}
i <- i + j + k
\end{sparks}
  \tcblower
\begin{asm}
. set |_tmp1| addw i j
. set |_tmp2| copyw k
. set i addw |_tmp1 _tmp2|
\end{asm}
\end{spasm}
\begin{spasm}
\begin{sparks}
i <- i + j + k + j
\end{sparks}
  \tcblower
\begin{asm}
. set |_tmp1| addw i j
. set |_tmp2| addw k j
. set i addw |_tmp1 _tmp2|
\end{asm}
\end{spasm}
\begin{spasm}
\begin{sparks}
z <-(x and y) 
      or (x and z)
\end{sparks}
  \tcblower
\begin{asm}
. set |_tmp1| and x y
. set |_tmp2| and x z
. set z or |_tmp1 _tmp2|
\end{asm}
\end{spasm}

\begin{spasm}
\begin{sparks}
z<-(A[0] and A[1]) 
   or (A[2] and A[3])
\end{sparks}
\tcblower
\begin{asm}
. set |_tmp1| array_ref A 0
. set |_tmp2| array_ref A 1
. set |_tmp3| and |_tmp1 _tmp2|
. set |_tmp4| array_ref A 2
. set |_tmp5| array_ref A 3
. set |_tmp6| and |_tmp4 _tmp5|
. set z or |_tmp3 _tmp6|
\end{asm}
\end{spasm}

\subsection{\texttt{if} blocks}
\change{The steps required for \Sparks{} control structures can be calculated
in terms of the steps required for the controlling expression(s) and
for those the body. The simplest case is the one-branched if (i.e. no
\lstinline[language=sparks]{else}).
\begin{equation*}
  \steps(\codevar{if}) \leq \steps(\Body) + \steps(\BoolExpr) + 2
\end{equation*}
}
\label{sec:orgcedd1e2}
\begin{spasm}
\begin{sparks}
if `\BoolExpr` then
  `\Body`
endif
\end{sparks}
  \tcblower
\begin{asm}
. set guard0 `\BoolExpr`
. unless guard0 else0
  `\Body`
else0 `\ldots`
\end{asm}
\end{spasm}

\subsection{\texttt{if-then-else} blocks}

\change{The \lstinline[language=sparks]{if-then-else} block is similar, except there
are two bodies to account for, and one more overhead step.}
\begin{equation*}
  \steps(\codevar{if-else}) \leq \max \left (\steps(\Body_t),\steps(\Body_f)\right) + \steps(\BoolExpr) + 3
\end{equation*}

\label{sec:orgcedd1e2}
\begin{spasm}
\begin{sparks}
if `\BoolExpr` then
  `$\Body_t$`
else
  `$\Body_f$`
endif
\end{sparks}
  \tcblower
\begin{asm}
. set guard0 `\BoolExpr`
. unless guard0 else0
  `$\Body_t$`
. goto done0
else0 `$\Body_f$`
done0 `\ldots`
\end{asm}
\end{spasm}

\subsection{\texttt{while} loops}

\change{
A while loop can be analysed in essentially the same was as an
\lstinline[language=sparks]{if} block, with the both the controlling
expression and the body executed one per iteration. In general (e.g.\
with nested loops) the cost of executing the loop body can vary
between iterations: we use the notation $\steps(\Body;i)$ to denote
the number of steps to execute the body on the $i$-th iteration. The
last execution of the controlling expression must be counted
separately.
}
\begin{align*}
  \steps(\codevar{while}) & =   \sum_{i \in \mathrm{iterations}}\left(\steps(\BoolExpr)+\steps(\Body;i)+3\right) \\
  & \qquad +  \steps(\BoolExpr)+2\\
\end{align*}

\begin{spasm}
\begin{sparks}
while `\BoolExpr`
      `\Body`
done
\end{sparks}
  \tcblower
\begin{asm}
while1 set _tmp `\BoolExpr`
.   set _test1 not _tmp
.   unless _test1 done1
.   `\Body`
.   goto while1
done1 `\ldots`
\end{asm}
\end{spasm}

\subsection{\texttt{for} loops}

\change{
\Sparks{} \lstinline[language=sparks]{for} loops have two integer
expressions defining the loop bounds, but these are only evaluated
once, outside the loop. As with \lstinline[language=sparks]{while}
loops, a careful analysis may need to distinguish between the costs of
different iterations, so we again use the notation $\steps(\Body;i)$
for the cost of the $i$th iteration of the loop body.
}
\begin{equation*}
  \steps(\codevar{for}) = \steps(\Lower) + \steps(\Upper)
  + \sum_{i=\Lower}^{\Upper}\left(\steps(\Body;i)+4\right)-2
\end{equation*}
\begin{spasm}
\begin{sparks}
for i<-`\Lower`,`\Upper` do
    `\Body`
done
\end{sparks}
\tcblower
\begin{asm}
. set i `\Lower`
. set _stop0 `\Upper`
for0 `\Body`
.    set _test0 eqw i _stop0
.    if _test0 done0
.    set i incw i
.    goto for0
done0 `\dots`
\end{asm}
\end{spasm}

Examples of \Sparks~code are given in Sections \ref{makespan} and \ref{matching}
for the makespan and maximum matching problem respectively.

\section{Linear programs from \Sparks}
  \label{sect:general}

The translation of a \Sparks~ program into an LP is adapted from a proof of
Cook's theorem given in \cite{HS78} which is attributed to Sartaj Sahni.
In Sahni's construction the underlying algorithm may be non-deterministic, but we
will consider only deterministic algorithms. Furthermore, Sahni describes how
to convert his pseudocode into a satisfiability expression. Although it would be
possible to convert this expression into an LP, considerable simplifications
are obtained by doing a direct conversion from the assembly code to an LP.
We give a brief overview of the LP variables and how
a simple assignment statement is implemented in inequalities in this section.
Full details of this conversion along with  sets of inequalities for the basic operations in \Sparks~were developed in
\cite{ABTW19}. Refinements were added during the implementation of the \Sparktope~compiler and are described in the
documentation at \cite{AB17a}.

The variables in the LP are denoted as follows. They
correspond to the values of variables in the source code at a given time $t$
in the execution.
\begin{itemize}
\item
{\em Binary variables}  $B(i,t), 1 \leqslant i \leqslant q(n), 0 \leqslant t < p(n)$. \\
$B(i,t)$ represents the value of binary variable $i$ after $t$ steps of computation.
For convenience we may group $W$ consecutive bits together
as an (unsigned) {\em integer variable} $i$. $I(i,j,t)$ represents the value of the $j$-th bit of integer
variable $i$ after $t$ steps of computation. The bits are numbered from
right (least significant) to left (most significant), the rightmost bit being numbered 1.
\item
{\em Binary arrays} A binary array $R[m],m=0,1,...,u$ is stored in consecutive
binary variables $B(\alpha + m,t), 0 \leqslant m \leqslant u, 0 \leqslant t \leqslant p(n)$ from
some base location $\alpha$. The array index $m$
is stored as a $W$-bit integer $I(*,*,t)$ and so we must have $u \leqslant 2^W -1$.
\item {\em 2-dimensional binary arrays}
A two dimensional binary array $R[m][c]$, $m=0,1,...,u$, $c=0,1,...,v$
is stored in row major order in consecutive binary variables
$B[\alpha+j,t-1]$, $0 \leqslant j \leqslant uv+u+v$,  $0 \leqslant t \leqslant p(n)$ from
some base location $\alpha$. The array indices $m$ and $c$
are stored as $W$-bit integers $I(*,*,t)$ and so we must have $u,v \leqslant 2^W -1$.  If $v=W$, the rows of $R[m][c]$ may be addressed as $W$-bit integers.

\item
{\em Step counter}  $S(i,t), 1 \leqslant i \leqslant l, 1 \leqslant t \leqslant p(n)$.\\
Variable $S(i,t)$ represents the instruction to be executed
at time $t$. It takes value 1 if line $i$ of the assembly code is being executed at time $t$
and $0$ otherwise.
\end{itemize}

All of the above variables are specifically bounded to be between zero
and one in the LP.  The last set of variables $S(i,t)$ encode the step
counter discussed in Section \ref{stepcounter}, and are crucial to the
correctness of our LPs.  The step to be executed at any time $t$ will
usually depend on the actual input.  For each time step $t$ and \change{\Asm{}
statement $i$ we will develop a system of inequalities which have the
{\em controlled x-0/1 property}.  This means they have the $x$-0/1
property (i.e.\ $0/1$ values of $x$ force unique $0/1$ values of the
remaining variables) for some subset of variables $x$, {\em if line
  $i$ is executed at time $t$}.  If line $i$ is not executed at time
$t$ the system of inequalities for line $i$ and time $t$ is
effectively redundant; more precisely the system is feasible for any
vector in the corresponding $[0,1]$ hypercube.  Each such set of
inequalities uses $S(i,t)$ as a control variable (i.e. they are
enabled if $S(i,t) = 1$).  This is just the usual ``big-M''
method where the value of $M=1$.}

As an example we consider a set of constraints that corresponds to the assignment
\lstinline[language=sparks]{s <- x xor y}.
Assume that $x,y,s$ are stored in $B(q,t-1), B(r,t-1), B(s,t)$ respectively.
\begin{eqnarray*}
S(i,t)+B(q,t-1)-B(r,t-1)-B(s,t) &\leqslant& 1~~~~~  \\
S(i,t)-B(q,t-1)-B(r,t-1)+B(s,t) &\leqslant& 1~~~~~  \\
S(i,t)-B(q,t-1)+B(r,t-1)-B(s,t) &\leqslant& 1~~~~~  \\
S(i,t)+B(q,t-1)+B(r,t-1)+B(s,t) &\leqslant& 3~~~~~
\end{eqnarray*}
If $S(i,t)=1$ then all constants on the right hand side are reduced by one and
$S(i,t)$ can be deleted. It is easy to check the inequalities have the
controlled $\{B(q,t-1),B(r,t-1)\}$-0/1 property, and that for each such 0/1 assignment $B(s,t)$
is correctly set.

We observe that each line of assembler code generates a set of constraints for each time step
of the run. In many cases there may be segments of code that either cannot run after a given
time step or cannot run before a given time step. A typical example of this is initialization
code that is run once at the beginning and never repeated. To reduce the total number of constraints
generated we introduced a compiler directed \texttt{phase} command. The user is required to supply
in the parameter file
a lower bound on the start time and upper bound on the finish time for each phase. 
Constraints are only generated for time steps falling inside this range.
We give an example of this for the matching problem discussed in Section \ref{matching}.

In order to create and solve an LP from an algorithm using \Sparktope~three steps are required.
A detailed description is given in the user's manual available at \cite{AB17a} and we give
only a summary here.
First the algorithm is written in \Sparks~and a parameter file is created to define the set of
instances to be solved. As a minimum this file includes the instance size, generically denoted $n$
in this paper, the number of bits $W$ required in the computation and an upper bound on the number
of computational steps required to solve any instance of size $n$. Using the \Sparks~code
and the parameter file an LP constraint set is computed for the family of inputs of size $n$. As a convenience for the programmer, arbitrary arithmetic expressions involving only parameters may be evaluated at compile time by enclosing them in \$\$.

The second step involves adding an objective function to this set of constraints to represent
the given instance to be solved. 
It is important to note that the constraint set is not rebuilt for each input instance 
of a given size.
A single constraint set is sufficient for all instances of size $n$.

Finally in the third step an LP solver is used
to compute the optimum solution of the LP and hence solve the given instance. Currently
\glpsol{} (default) and \cplex~are supported directly.
\change{Due to variations in the LP format accepted by different solvers, we converted the compiler output to MPS before
loading it with \gurobi{}.\footnote{We learned \cite{MM2020} while the paper was under revision that with small changes the LP files can be read directly by Gurobi. Unfortunately this does not improve the results.}}
 The LPs produced by our compiler appear to be
quite difficult to solve with numerical problems often encountered.
For this reason an option is given to check the given LP solution using exact arithmetic (currently only supported using \glpsol{}).
Also to speed the solution we have a {\em -f} option. In this case at the second step
described above
in addition to inserting the objective function the $n$ input variables
are set to the 0/1 values corresponding to the given instance. 
As proven in Proposition \ref{aboutd} these
are their values at optimality. The job of the LP solver is to find the 0/1 values of the other
variables which in turn give a full trace of the \Sparks~code on the given instance
and hence the solution for the given instance.
The LPs produced have an extremely large number of variables and this can make debugging
the original \Sparks~code into a challenge. To assist in this process we provide 
a visualization of the entire run of the code based on the output of the LP solver.

In the next two sections we give two examples of algorithms converted to LPs by \Sparktope.
The first is a simple optimization problem called makespan and the second is a decision
problem to determine if a matching in a graph is maximum or not.
\change{
In discussing the complexity of an LP we follow the usual practice in
mixed integer programming of dropping the dependence on the word size $w$
and define it to be the number of constraints plus number of variables.
}

\section{Makespan}
\label{makespan}
The makespan problem is to schedule $m$ jobs on $n$ identical machines to minimize the finishing
time, known as the makespan, of the set of jobs. 

\change{
\begin{description}
\item[Instance] Integers $m$,$n$, and job times $p[0],\dots ,p[m-1]$.
\item[Problem] Schedule $m$ jobs on $n$ machines to minimize the latest finish
         time (makespan) $T$.
\item[Output]  A job schedule $x[m,n]$ where $x[i,j]=1$ if job $i$ is scheduled on
         machine $j$ and zero otherwise.
\end{description}
}

\begin{figure}
  \centering
    \includegraphics[height=0.97\textheight]{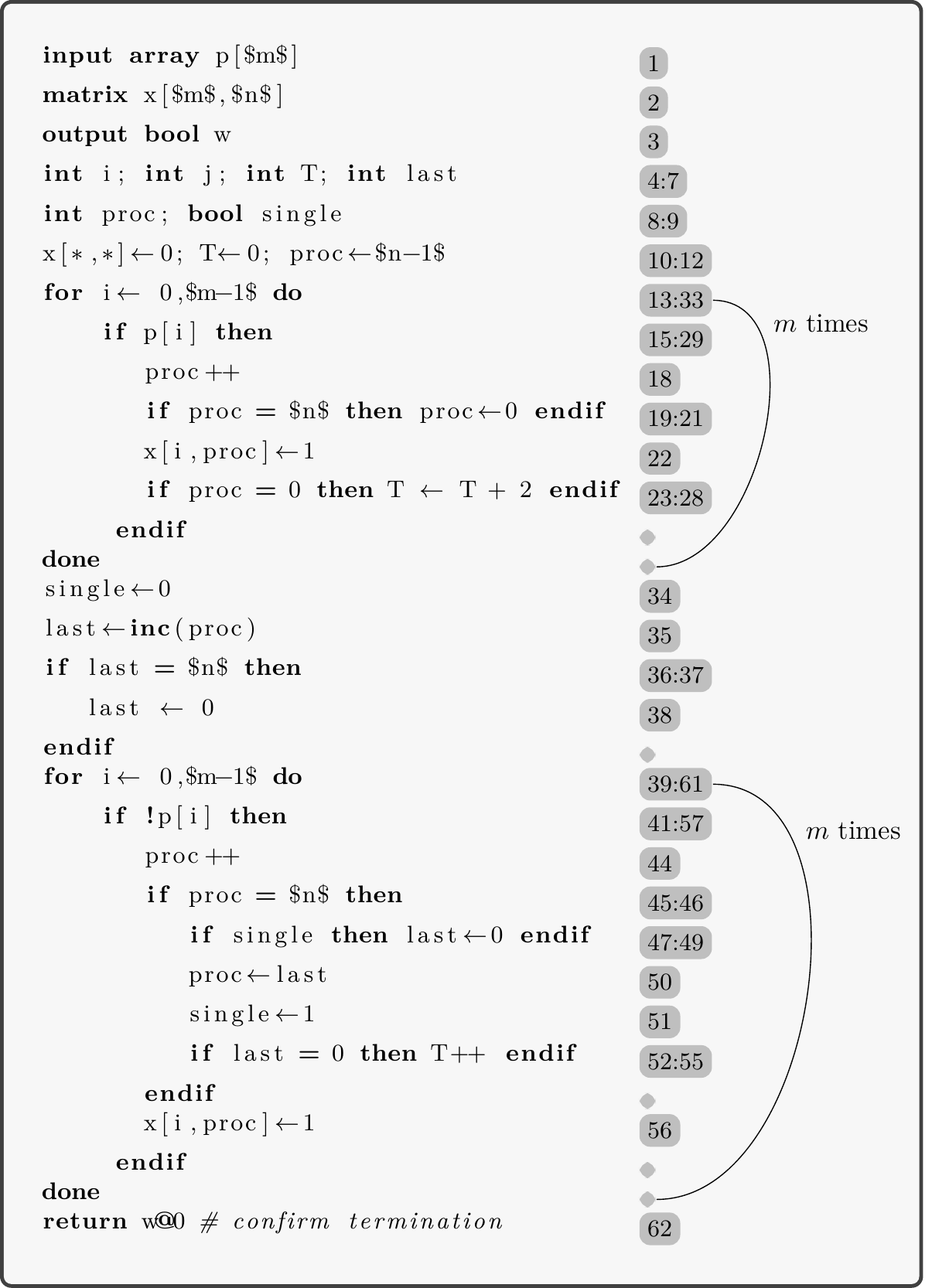}
\caption{\Sparks{} source code {\em ms.spk} }
  \label{sparks}
\end{figure}

This problem has
exponential extension complexity even when the
job times are in $\{1,2\}$ and $T=2$, see Tiwary, Verdugo and Wiese \cite{TVW19}.
We restrict ourselves to this case. It is easily seen that a greedy
algorithm that schedules all the jobs with processing time 2 first
and then the remaining jobs with time 1 gives the optimum schedule.
This is achieved by the code \texttt{ms.spk} given in Figure~\ref{sparks}.
The right hand column gives the line numbers of the assembler code
instructions that are generated from the source code following
the descriptions given in Section \ref{stepcounter}.
Due to space limitations we do not give the assembly code here but it
is available at \cite{AB17a}.
Since the processor times are either 1 or 2 the
input for \texttt{ms.spk} can be given by specifying a boolean array
\texttt{p} where the processing time of job $i$ is $p[i]+1,~i=1,2,...,n$. 
The variable $w$ confirms program termination, mainly
for debugging.

As written, the code requires $O(nm)$ space to hold the schedule $x$
and $O(m)$ time as it consists of two unnested \texttt{for} loops each
run $m$ times.
It follows from results in \cite{ABTW19} that the LP produced has complexity \change{(i.e.\ inequalities and variables)} $O(nm^2)$.
Another implementation could use an integer array $x$ of length $m$ where
$x[i]$ is the processor assigned to job $i$. Assuming $m \ge n$ the word size for integers is
$O(\log m)$ and so $O(m \log m )$ space is required for $x$ and the resulting
LP has complexity $O(m^2 \log m)$.
To actually produce linear programs for this problem we need to bound,
for each $m$, 
the maximum number of assembler steps taken for any instance of this size.
We do this below.
For a given integer $m$ we will give an upper bound
on the number of assembler code steps in order to reach a
return statement. For this discussion we will refer to
Figure~\ref{sparks}, where the range of assembly lines corresponding to
each \Sparks{} line is given. 

The first 9 lines are variable declarations and are not executed, so
there are 53 lines of executable \Asm{} code.  We see there are two
unnested \texttt{for} loops each executed $m$ times.  The first has 19 lines
(lines 15-33) and the second has 21 lines (lines 41-61).  So an upper
bound on number of steps executed is $53+40(m-1)=40m+13$. However
inspecting the two \texttt{for} loops we see that they have mutually exclusive
\texttt{if} statements depending on the value of the input $p[i]$.  The body of
these \texttt{if} statements are contained between lines 18-28 and 44-55,
respectively, and only one of these blocks can be executed for each
$i$. The shorter first block contains 11 statements so we may reduce
the overall running time by $11m$ obtaining the upper bound $29m+13$
for the number of assembler steps executed.  A slightly tighter
analysis is possible by observing that some of the assembly statements
for a \texttt{for} loop are executed only once.

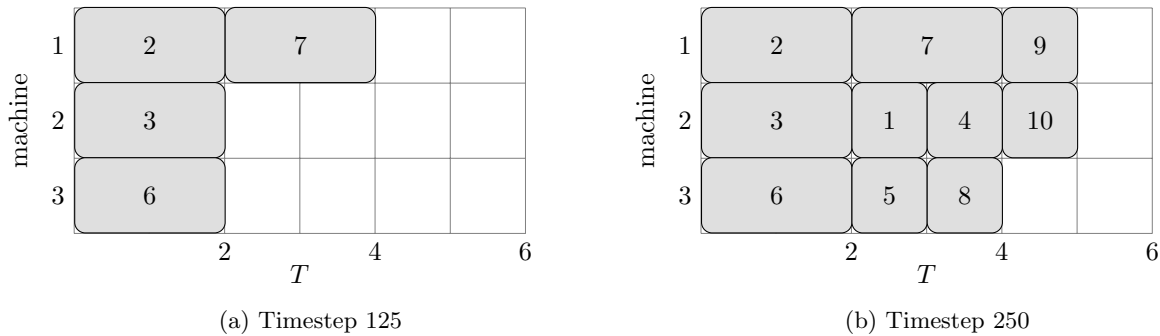
\begin{figure}[!ht]
\newcommand{\job}[4]{\draw (#1,#2) node [draw,fill=gray!25,minimum width=#3cm,minimum height=1cm,rounded corners,anchor=south west]  {#4};}

\newcommand{\mlabels}{
    \draw (-0.5,1.5) node[rotate=90,anchor=south] {machine};
    \draw (0,1) node[anchor=south east,minimum height=1cm] {1};
    \draw (0,2) node[anchor=south east,minimum height=1cm] {2};
    \draw (0,3) node[anchor=south east,minimum height=1cm] {3};
}
\newcommand{\tlabels}{
    \draw (2,3) node[anchor=north] {2};
    \draw (4,3) node[anchor=north] {4};
    \draw (6,3) node[anchor=north] {6};
    \draw (3,3.3) node[anchor=north] {$T$};
}
  \begin{subfigure}{0.5\textwidth}
    \begin{tikzpicture}[yscale=-1]
      \draw[gray,very thin](0,0) grid  (6,3);
      \job{0}{1}{2}{2}
      \job{0}{2}{2}{3}
      \job{0}{3}{2}{6}
      \job{2}{1}{2}{7}
      \mlabels
      \tlabels
    \end{tikzpicture}
    \caption{Timestep $125$}
  \end{subfigure}
  \hfill
  \begin{subfigure}{0.5\textwidth}
    \begin{tikzpicture}[yscale=-1]
      \draw[gray,very thin](0,0) grid  (6,3);
      \job{0}{1}{2}{2}
      \job{0}{2}{2}{3}
      \job{0}{3}{2}{6}
      \job{2}{1}{2}{7}
      \job{2}{2}{1}{1}
      \job{3}{2}{1}{4}
      \job{2}{3}{1}{5}
      \job{3}{3}{1}{8}
      \job{4}{1}{1}{9}
      \job{4}{2}{1}{10}
      \mlabels
      \tlabels
    \end{tikzpicture}
    \caption{Timestep $250$}
  \end{subfigure}
  \caption{Greedy construction of schedule from input \texttt{ms10}}
  \label{fig:schedule}
\end{figure}

\begin{figure}[!ht]
  \centering
    \includegraphics[width=0.9\textwidth]{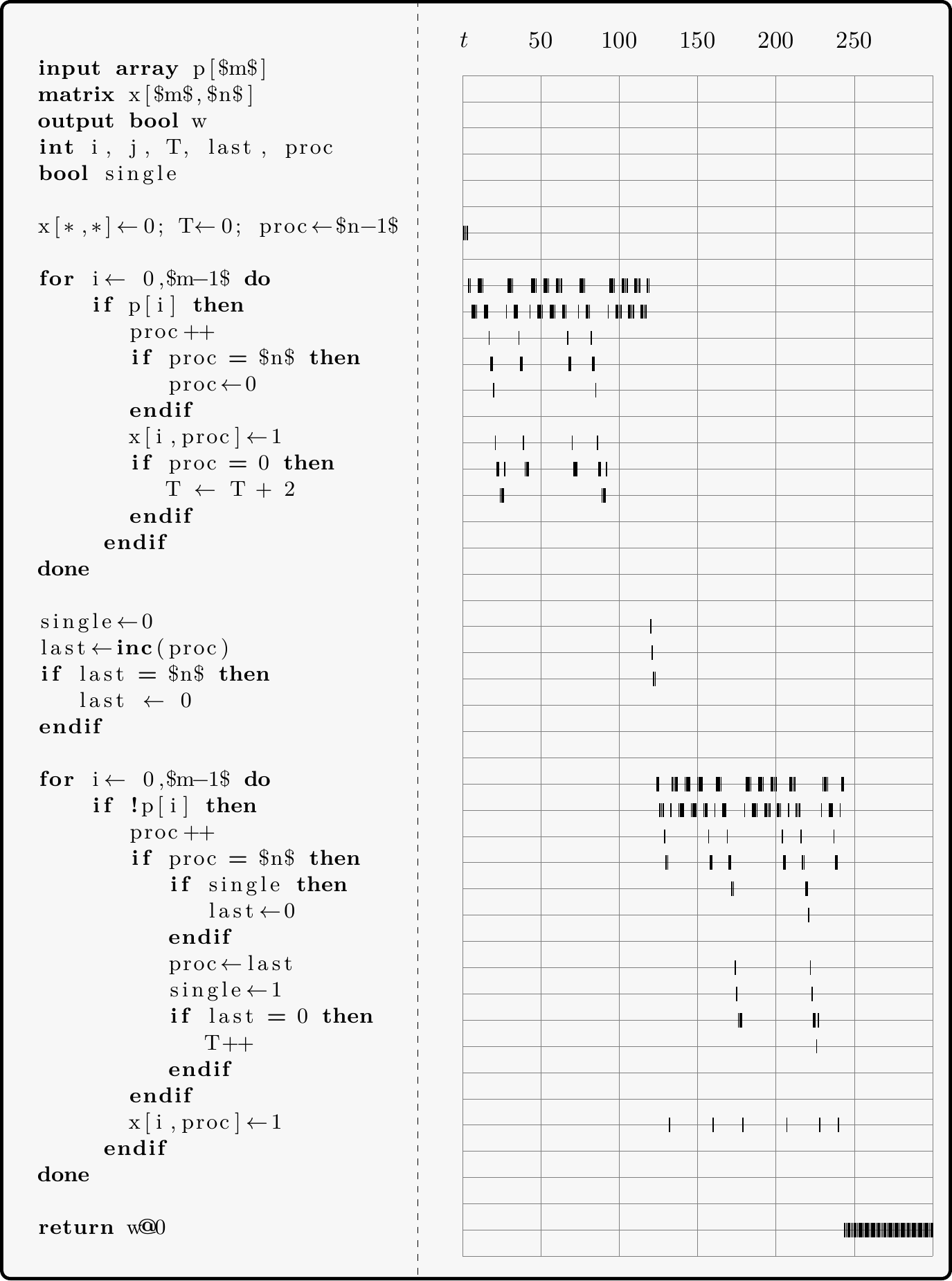}
\caption{Trace of run on input \texttt{ms10}  }
  \label{ms10}
\end{figure}

An example input \texttt{ms10} is provided for the case $m=10, n=3$.
The \Sparks{}
input is:
\begin{verbatim}
                  array p[10] <- {0, 1, 1, 0, 0, 1, 1, 0, 0, 0}
\end{verbatim}

The list decreasing algorithm first schedules the jobs with processing time 2, as shown
on the left in Figure~\ref{fig:schedule}.
The optimal schedule is shown on the right and has makespan $T=5$.
A complete trace of the run is given in
Figure~\ref{ms10}. It shows which line is executed at each time step.
This is achieved by observing in the LP optimum solution for each time $t$
the unique value of $i$ for which $S(i,t)$=1.
We see that the run terminated at around time $t=245$ at the return statement 
in the last line of code.
This termination time compares with
the upper bound of $29*10+13=303$ steps calculated above.
In the trace it can clearly be seen how at around $t=125$
the code switches from scheduling the jobs with processing time
2 to those with time 1.

\begin{table}
  \tbl{\change{Makespan problem sizes, run times in seconds, computed makespan and steps to compute}.  (secs) indicates non 0/1 solution}
  {\begin{tabular}{|c|c|c|c|c|c|c||c|c|c||c|c|}
     \hline
     \multicolumn{7}{|c||}{Linear Programs}   &\multicolumn{3}{c||}{Solvers} &\multicolumn{2}{c|}{Outputs} \\
     name &m &max&rows &columns&non-zeros& file& \glpsol  & \cplex & \gurobi & T & steps \\
     & & steps& ($\times{}1000$)& ($\times 1000$) & ($\times 1000$) & MB&  &   & &  &  \\
     \hline
     \texttt{ms5}& 5 &201 &  \change{129} &\change{28} &43& 7 & 0.1 & 0.1 & 0.3 & 3   & 134   \\
     \texttt{ms10}&10  &321 &380 &59& 1,305&21 & 0.4 & 0.3 & (31)  & 5 &244 \\
     \texttt{ms20} &20 &631 &1,170 & 149 & 4,160&66 & 1.2 &0.8 & (91)  &10 &483 \\
     \texttt{ms40} &40 &1251 & 3,845 & 412 &14,099& 226 & 3.9 & 2.8 & (107)  &20 &958 \\
     \texttt{ms80} &80 &2491 & 13,483 & 1,251 & 50,635&833 & 15 & 10 & (587)  &39 &1899\\
     \texttt{ms160} &160 &4971 & 49,869 &4,151  &190,407& 3287 & 55 & 61 & (4422)  &78 &3790\\
     \hline
   \end{tabular}}
\label{tab:ms}
\end{table}

Table \ref{tab:ms} shows some test results on makespan problems for
$m=5,10,20,40,80,160$ with $n=3$\footnote{All runs on \texttt{mai20}: 2x Xeon E5-2690 (10-core 3.0GHz), 20 cores, 128GB memory, 3TB hard drive}. The corresponding LPs (solver inputs) are available from~\cite{AB20a}.

The first 6 columns describe the linear programs generated for various values of $m$,
the number of processes. They give the number of rows and columns in each LP
and its size in MB using \cplex~LP format. The step bound is the upper bound 
of $31m +11$ on
the number of steps required to reach a return statement, computed above.
Therefore these LPs will solve any instance of the given size $m$ by simply
changing the objective function.
As $m$ doubles we can see the input size goes up
a little less than 4 times, indicating the predicted quadratic behaviour
(since $n$ was held constant).
We used three solvers, \glpsol~4.55, \cplex~12.6.3 and \gurobi~8.1.1~each with
default settings. In order to use \gurobi~we converted all files to the larger
MPS format using \glpsol. Without fixing the
input variables to their given 0/1 values via the {\em -f}
option described earlier, all solvers gave fractional non-optimal
solutions. With the input variables fixed \glpsol~and \cplex~correctly
found the optimum 0/1 solution in each case using \change{preprocessing}.
However \gurobi~could only find a 0/1 solution for $m=5$ and found fractional
solutions for the other cases. The final two columns describe the outputs
obtained on a test problem for each LP:
the makespan $T$ and the actual number of steps executed to reach the return statement.

\lstset{language=sparks,frame=single}

\section{Maximum matching}
\label{matching}
We consider the following matching problem:

\change{
  \begin{description}
  \item[Instance] Integer $n$, graph $G$ and matching $M$ in $G$.
  \item[Problem]  Decide whether $M$ is a maximum matching, and if it is not,
         to find an augmenting path.
  \item[Output]  $w=0$ if $M$ is a maximum matching
         $w=1$ if there is an augmenting path
  \end{description}
}
This standard formulation of this problem has exponential extension complexity, as follows from the
result of Rothvoss \cite{Rothvoss17}.
However, one iteration of Edmonds' blossom algorithm \cite{Edmonds1965a} can answer the above  problem
and this is achieved by the code \texttt{mm.spk} given in Appendix \ref{spkcode}.
It is based on the pseudocode in \cite{wikiBlossom} and a detailed explanation
and proof of correctness is given in Section 16.5 of Bondy and Murty \cite{BM2008}.
Due to space limitations we do not give the assembly code \texttt{mm.asm} here but it
is available at \cite{AB17a}. As with the Makespan example we have added the
relevant assembly code line numbers to the \Sparks~code.

To produce linear programs for this problem we need to bound,
for each $n$, 
the maximum number of assembler steps taken for any instance size $n$.
We do this in the next section getting an upper bound of less than $31n^3$ steps.
Since the space required is $O(n^2 \log n)$
it follows from results in \cite{ABTW19} that the LP produced has $O(n^5 \log n)$
constraints.

\subsection{Step count analysis}

For a given integer $n$ we will give an upper bound on the number of steps taken
in order to reach a return statement. Referring to the source code
 \texttt{mm.spk} we see that the program is divided into two phases to reduce the number
of constraints produced.
\texttt{phase} {\em init} handles initialization and is executed once
whereas \texttt{phase} {\em main} performs the blossom algorithm.

The loop structure of \texttt{phase} {\em init} is shown
in Figure \ref{init}. The line numbers correspond to the assembly code 
\texttt{mm.asm} and as the first 28 lines are declarations they are omitted. 
It is necessary to get both an upper and lower bound on the number of steps 
executed in \texttt{phase} {\em init}.
A total of 7 lines are executed once: 29-32, 38-39, and 59.
The first \texttt{for} loop in lines 33-38 has 6 steps but the \texttt{done} statement is only executed once
and steps 36 and 37 are executed $n-1$ times.
So this loop requires exactly $5n-1$ steps.
For simplicity in what follows a \texttt{for} loop with $k+1$ steps, including the \texttt{done}
statement,
is assigned an upper bound of $kn$ steps and a lower bound of
$(k-1)n$ steps. 

There remain two nested \texttt{for} loops in lines 20-59. The inner \texttt{for} loop in lines
42-54 is executed $n-1$ times with the corresponding number of iterations
$=n-1,n-2,...,1$.  Since the loop has 13 lines an upper bound on the
number of steps executed is therefore
$12(n-1+n-2+...+1)=6n(n-1)$.
The remaining 7 lines in the outer \texttt{for} loop
40-58 contribute at most $7n$ steps. In total we have an upper
bound of $7+6n^2-6n+7n=6n^2+n+1$ for \texttt{phase} {\em init}. 

For the lower bound we note that 
lines 47,48 are executed only for edges
in an input matching and there may be zero of those. So we reduce the size of
the inner loop by 3 getting a lower bound of $9n(n-1)/2$ steps.
For the outer loop we reduce its size to 6 so in total the lower bound is
$7+9n(n-1)/2+6n=(9n^2+3n+14)/2$ steps.

For $n=8$ we have an upper bound of 393 and a lower bound of 307 steps. Referring to the
output snapshot in Figure \ref{wt8} we see that 
S[59,364]=1 and
S[62,365]=1
so that for this input 364 steps were executed in \texttt{phase} {\em init}.

\begin{figure}

\begin{sparksbox}{basicstyle=\scriptsize}
     phase init do
29     
     ...one time assignments...
32   
33   for i<- 0,$n-1$ do
          .........
38   done
39   ...one time assignment...
40   for i<- 0,$n-2$ do
        .........
42	for j<-inc(i),$n-1$
          .........
          if a[j,i] then   # matching edge: j  i
47            match[[j]] <- i
48            match[[i]] <- j
          endif

54      done
        .........
59   done
     done                  # phase init
\end{sparksbox}

\caption{Control structure of phase init of \texttt{mm.spk} }
\label{init}
\end{figure}

\begin{figure}
\begin{sparksbox}{basicstyle=\scriptsize}
     phase main do
62   while progress do      #Loop A: find aug path and exit or find blossom and shrink
         ...
68       for i<-0,$n-1$ do
            ...
82       done
         ...
85       while !progress and !doneV do         #Loop B: process unexplored vertex V
         .      ...
100      .      while !progress and !doneW do  #Loop C: process unmarked edge VW
         .      .     ...
124      .      .   if  !F[W] then   # add edge to F
         .      .       ...
133      .      .   else             
                     .
                     .               # see Figure below
                     .
239      .      .   endif            
         .      .       ...
247      .      done                 # end of Loop C
         .      ...
255      done                        # end of Loop B     
257  done                            # end of Loop A  
258  return w @ 0                    # no augmenting matching
260  done                            # phase main
\end{sparksbox}
\caption{Control structure of phase main }
\label{main}
\end{figure}

\begin{figure}
\begin{sparksbox}{basicstyle=\scriptsize}
133  else
134        while i != parent[[i]] do    # Loop D
             ...
140        done
           ...
142        while k != parent[[k]] do    # Loop E
              ...
148        done
              ...
151        if  i != k then       # augmenting path
               return w @ 1
154        else                  # shrink blossom
               ...
157            while V != X  do         # Loop F
                   ...
165            done
                   ...
169            while V != W do   #Loop G: traverse and shrink cycle
               .       ...
196            .       while j<V do         # Loop H
               .         ...
213            .       done
214            .       while j != $n-1$ do  # Loop I
               .         ...
233            .       done
               .       ...

236            done          # traverse and shrink cycle
238        endif             # if i != k
                ...
\end{sparksbox}
\caption{Control structure of path and blossom processing }
\label{shrink}
\end{figure}

We now turn to the main part of the blossom algorithm and give the control structure in
Figure \ref{main}. There are three nested \texttt{while} loops labelled A,B,C respectively. 
The outer Loop A, lines
62-257, finds either an augmenting path or a blossom. 
A blossom is an odd cycle of
length at least three which is shrunk to a single vertex, removing at least 2 nodes
of the graph. Graphs cannot be shrunk to less than 3 vertices,
so shrinking can happen at most $(n-3)/2$ times and this is a bound on the
number of times Loop A can be executed. 
Loop B, lines 85-255 is executed
for each vertex in the (possibly shrunk) graph, so $n$ times at most
for each iteration of Loop A. 
Loop C, lines 100-247, is the third nested
loop, and is executed at most once for each vertex adjacent to the
vertex chosen in Loop B. 
So at most $n$ iterations are required for each iteration of Loop B.

We begin by analyzing the inner Loop C.
The block of code (line numbers 133-238) shown in Figure \ref{shrink}
either finds an augmenting path or handles blossom shrinking.
It can be executed at most once for each iteration of Loop A,
so $(n-3)/2$ times in total.
It is analyzed separately below. 
Removing these 106 lines from the 148 lines in Loop C
leaves 42 lines that require at most $42n$ time steps for each iteration of Loop B.
Moving to Loop B and we find 23 lines
(85-99, 248-255) which are not in Loop C.
Loop B executes $n$ times for each iteration of Loop A and so requires
at most $42n^2+23n$ time steps for such an iteration.

Finally in Loop A there is a \texttt{for} statement, lines 68-82, executed $n$ times for a total of
at most $15n$ steps per iteration.
There remain 10 lines (62-67, 83-84, 256-57)
executed once for each iteration.
So adding in the steps for the inner loops (except shrinking)
an iteration of Loop A requires at most $42n^2+38n+10$ steps. 

We now turn to the code in Figure \ref{shrink} which is executed at most $(n-3)/2$ times.
Consider a single iteration of these 106 lines.
Loops D and E of 7 lines each can be executed at most $n$ times each for a total of $14n$ steps.
Inside the \texttt{else} clause beginning on line 155
are several more loops. Loop F, lines 157-165, is executed at most $n$ times for
a total of $9n$ steps. 
Loop G, lines 169-236, is more complex. Consider a single iteration.
There are two further
\texttt{while} loops, Loop H on lines 196-213 and Loop I on lines 214-233. 
Since $j$ increments every time H of I runs, together these are executed at
most $n$ times in an iteration of Loop G.
The second of these is longer and has 20 lines.
So both loops together take at most $20n$ steps for each of at most $n$ iterations,
or $20n^2$ steps in total.
The remaining 30 lines (169-195, 234-236) of Loop G are executed once per iteration.
The total number of steps taken in Loop G for a single blossom shrinking is 
therefore at most $20n^2+30n$.
In this code block there remain lines 141, 149-156, 166-168, 238, or a total of 12 lines,
each of these is executed once per iteration of the code block.
It follows that the total number of steps taken in blossom shrinking is at most
$(14n+9n+20n^2+30n+12)=20n^2+53n+12$.

Putting everything together the total number of steps for an iteration of Loop A
is at most $62n^2+91n+22$. Since this loop executes at most $(n-3)/2$
times this gives an upper bound of $(62n^3-75n^2-251n-66)/2$ steps.
To this we add the upper bound on the steps taken in \texttt{phase} {\em init}, derived above,
of $6n^2+n+1$ obtaining a total of $(62n^3-63n^2-249n-64)/2$ steps.

\subsection{Examples}
\label{examples}

\begin{figure}[!ht]
\begin{subfigure}{.25\textwidth}
  \centering
  \includegraphics[width=.7\linewidth]{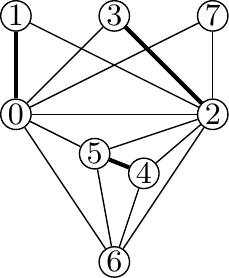}
  \caption{Graph and matching}
  \label{fig:sfig1}
\end{subfigure}%
\begin{subfigure}{.25\textwidth}
  \centering
  \includegraphics[width=.7\linewidth]{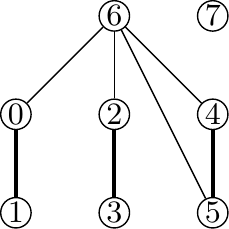}
\par\bigskip
\par\medskip
  \caption{Blossom 456 found}
  \label{fig:sfig2}
\end{subfigure}
\begin{subfigure}{.25\textwidth}
  \centering
\par\bigskip
  \includegraphics[width=.7\linewidth]{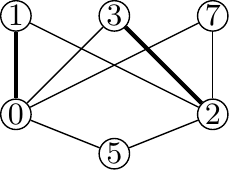}
\par\bigskip
\par\bigskip
  \caption{Blossom shrunk to 5}
  \label{fig:sfig1}
\end{subfigure}%
\begin{subfigure}{.25\textwidth}
  \centering
  \includegraphics[width=.7\linewidth]{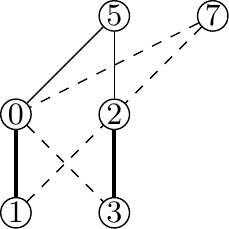}
\par\bigskip
  \caption{No augmenting path}
  \label{fig:sfigd}
\end{subfigure}

\caption{Processing the graph \texttt{wt8}}
\label{wt8_graphs}
\end{figure}

Two example inputs are provided in Appendix \ref{inputs} for the case $n=8$, \texttt{wt8} and \texttt{wt8a}.
The graphs have vertices labelled 0,1,...,7 and the input matching has 3 edges 
(0,1),(2,3),(4,5)
as shown for $wt8$ in Figure \ref{wt8_graphs} (a).
\change{LPs (solver inputs) discussed in this section can be downloaded from~\cite{AB20a}.}

\begin{figure}
  \centering
    \includegraphics[width=0.8\textwidth]{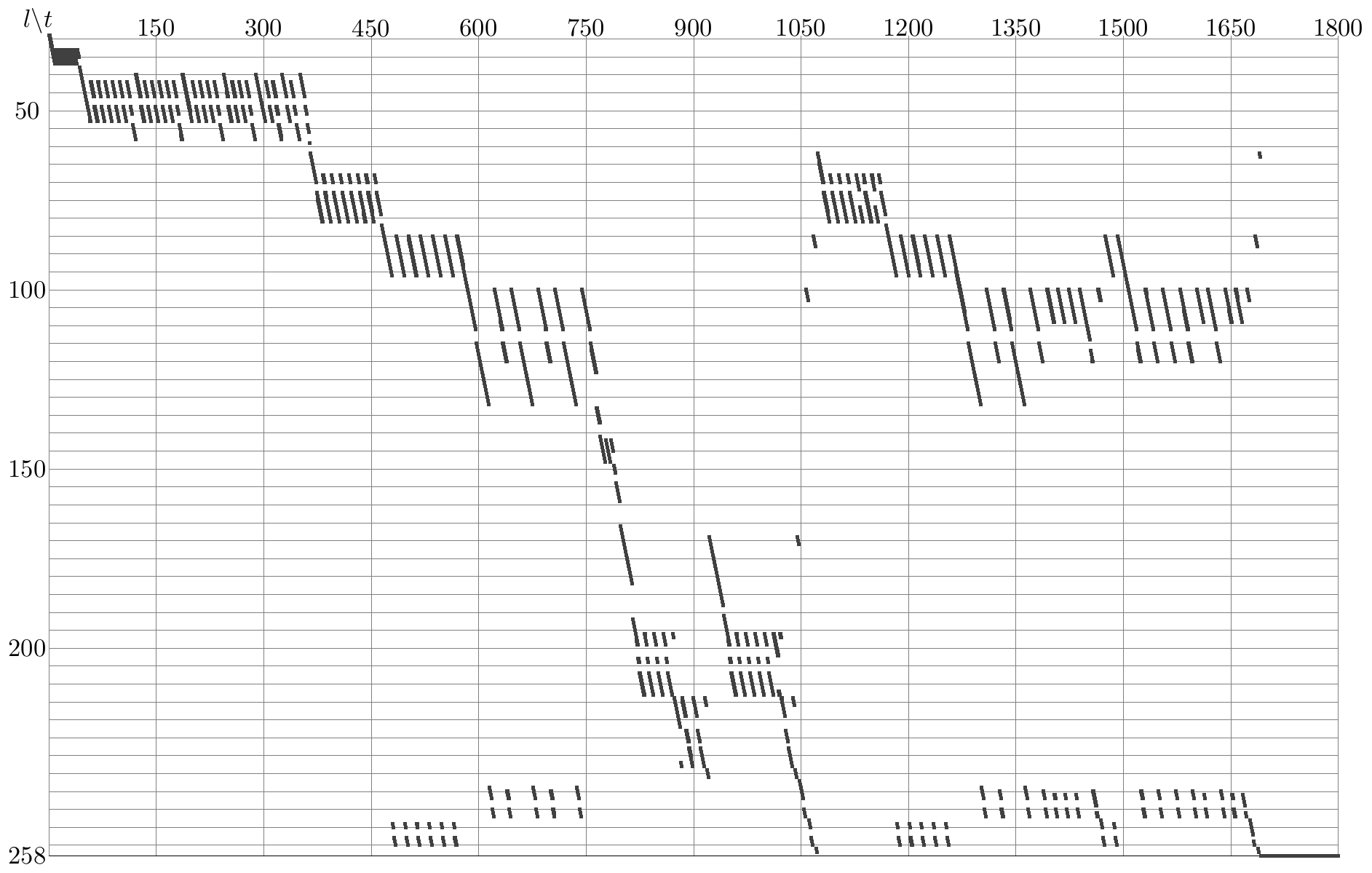}
\caption{\texttt{wt8}: no augmenting path}
\label{wt8}
\end{figure}

A snapshot of this run is given in Figure \ref{wt8}. We see that 
it terminates at a return statement on line 258 at around time 1700
indicating that an augmenting path was not found.
The step count compares with
the upper bound of 8123 steps calculated above.   

In Figure \ref{wt8_graphs}(b) we see how the algorithm first finds a blossom
on vertices 4,5,6 and shrinks it to vertex 5 
as in Figure \ref{wt8_graphs}(c). In the subsequent iteration no blossom
or augmenting path is found, as shown in 
Figure \ref{wt8_graphs}(d), to the run terminates.

\begin{figure}
  \centering
    \includegraphics[width=0.8\textwidth]{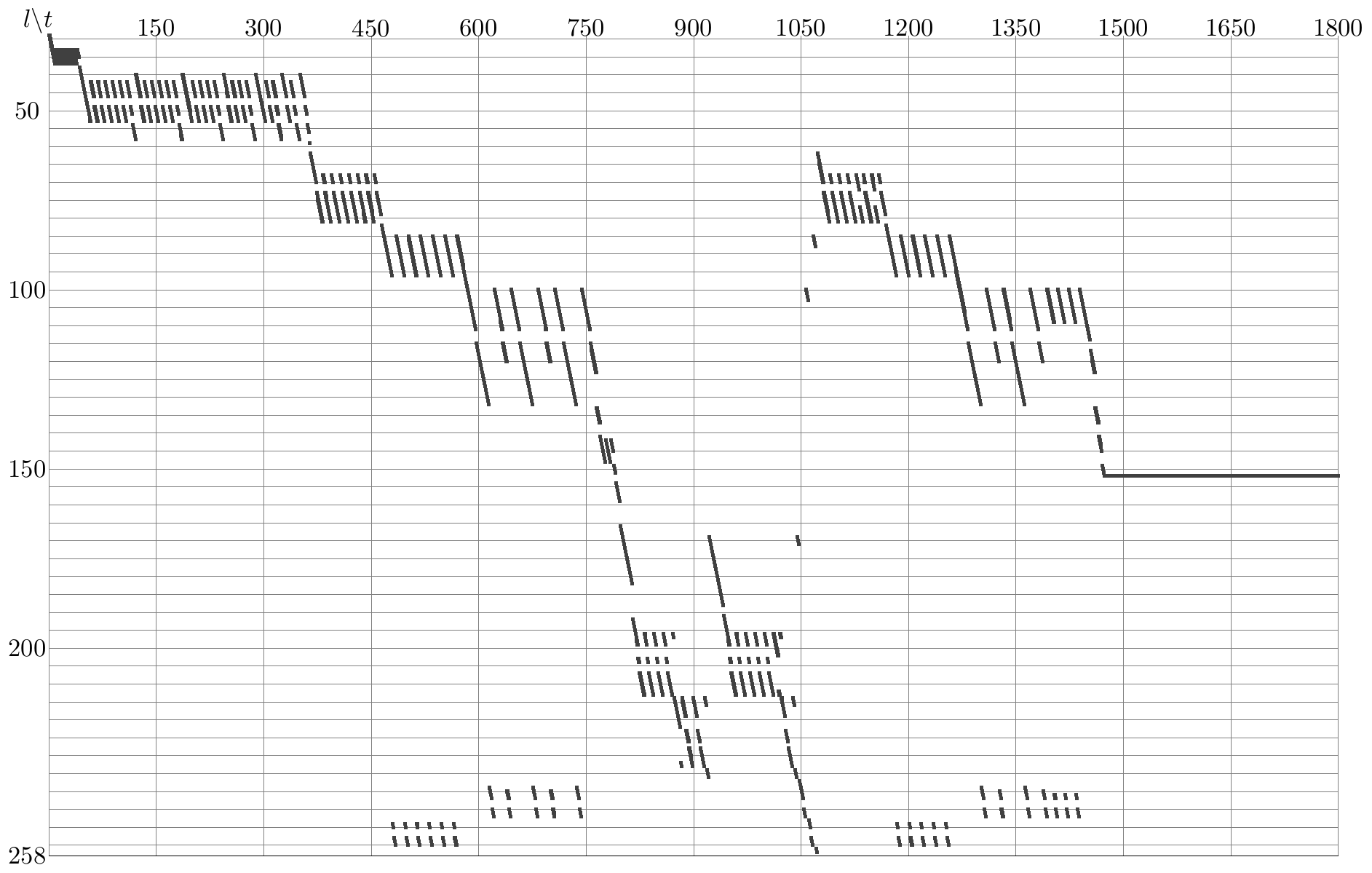}
\caption{\texttt{wt8a}: augmenting path found}
\label{fig:wt8a}
\end{figure}

The input  \texttt{wt8a} contains the additional edge 47. As before a blossom on vertices
4,5,6 is found and shrunk to vertex 5. However this time an augmenting path 57 is 
found in the shrunk graph. This expands to the augmenting path 6,5,4,7 in \texttt{wt8a}.
Observing the snapshot in Figure \ref{fig:wt8a} we see that 
the run halts on line 152 of the code after about 1480 steps and returns $w=1$.

\begin{table}[htb]
\tbl{Linear programs generated for maximum matching}
{\begin{tabular}{|c|c|c|c|c|c|c|c|c|}
\hline
name &n &max steps&main.LB&init.UB&rows&columns&non-zeros&GB\\
\hline
\texttt{mm8}& 8 &4000 (9747) & 307 & 393  &  \change{25,490,809}  &2,567,920& 80,568,489   &\change{1.6} (3.4)   \\
\texttt{mm10}&10& 7000 (19629)  &472 & 611   & 54,809,388 & 5,354,967  &  210,572,706   & 3.6 (11)\\
\texttt{mm12} &12 & 10000 (34771) &673 & 877   &94,860,776&8,200,011  & 371,213,800 &6.3 (23) \\
\texttt{mm16} &16 & 16000 (83003) &1183 & 1553  &  \change{238,577,463}  & \change{15,296,088} & \change{955,445,591} & 15 (80) \\
\hline
\end{tabular}}
\label{tab:mm}
\end{table}

\begin{table}[htb]
\begin{minipage}{0.64\textwidth}
\tbl{Maximum matching test results}
{\begin{tabular}{|c|c|c|c||c||c||c|c|}
\hline
\multicolumn{4}{|c||}{Inputs} &\multicolumn{1}{c||}{\glpsol}&\multicolumn{1}{c||}{\cplex} &\multicolumn{2}{c|}{Outputs} \\
name & n & m    & M &  secs   & secs& answer & steps \\
\hline
\texttt{wt8.in}& 8& 15  &  3  & 32  & 27   & max & 1692   \\
\texttt{wt8a.in}& 8 & 16 & 3      &  37   & 26   & aug & 1474\\
\hline
\texttt{wt10.in}& 10 &19& 4   &98    & 240  &max  & 1627 \\
\texttt{tr10.in}& 10 & 14  &4    & 97   & 272 & aug & 3733  \\
\hline
\texttt{wt12.in}& 12 & 24  & 5   &188& 319  &max  & 2671 \\
\texttt{tr12.in}& 12 & 17  &5    & 189  & 420 & aug & 5295  \\
\hline
\texttt{wt16.in}& 16 & 43  & 7  & - & 1353   &max     & 4241 \\
\texttt{tr16.in}& 16 & 23  & 7  & -    &1956   &aug    & 9211 \\
\hline
\end{tabular}
}
\label{tab:mm2}
\end{minipage}\hfill
\begin{minipage}{0.34\textwidth}
\centering
    \includegraphics[width=0.6\textwidth]{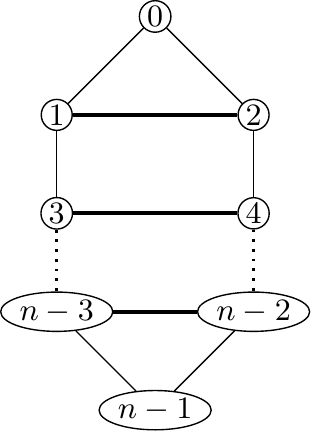}
\captionof{figure}{\texttt{trn.in}}
\label{trn}
\end{minipage}
\end{table}

Table \ref{tab:mm} shows some statistics about linear programs built for the maximum matching problem.
These are much bigger LPs than those generated for the makespan problem and the analysis
of the worst case run time given above is quite loose. To make it easier for
the solvers we generated LPs with
a time bound somewhat less than the proven worst case. 
The time bounds used are shown in column 3 with the worst case bound in parenthesis.
The \texttt{phase} bounds in columns 4 and 5 are as given above and the corresponding
statistics of the LP generated are given in the next 3 columns. The disk
space required to store the LP is given in the final column with the size of
the LP for the proven time bound in parentheses.

Table \ref{tab:mm2} shows the results of solving the LPs for some given input 
graphs\footnote{All runs on \texttt{mai32ef}: 4x Opteron 6376 (16-core 2.3GHz), 64 cores, 256GB memory, 4TB hard drive}.
Here $n$ denotes the number of vertices, $m$ the number of edges and $M$
the size of the given matching to test for being of maximum size.
The graphs \texttt{wtn.in} derive from Tutte's theorem and have no maximum matching,
so no augmenting path is found.
The graphs \texttt{trn.in} shown in Figure \ref{trn} achieve the maximum
number, $(n-2)/2$ of shrinkings, successively matching edges (1,2),(3,4),...,$n-3,n-4$,
before finding an augmenting path from vertex 0 to vertex $n-1$.
We observe that the steps used in finding the solution, shown in the final column,
are significantly smaller than the time bounds given in column 3 of Table \ref{tab:mm}.
\change{We used the same solvers and default settings as before, except \glpsol{} at version 4.60 (instead of 4.55).}
\glpsol~has a limit of $10^8$ constraints and so was not able to handle
\change{\texttt{mm16}}. Both \cplex~and \glpsol~solved all problems using the presolver only.
\gurobi~was not able to solve any of the models in the presolver and produced 
incorrect non integer solutions after pivoting.

\section{Concluding remarks}
\label{sect:conclusions}
\change{
Extension complexity initially was concerned with giving exponential lower
bounds for LP formulations of NP-hard problems that project onto their natural
formulations. The main examples being the travelling salesman problem
and the max-cut problem.
These results are independent of the $\Pp~\ne~\NP$ conjecture.
Later work showed that similar results also applied to some problems in $\Pp$.
The makespan and maximum matching problems are two examples of this.
Since we know that polynomial size LPs exist for these problems,
it is natural to wonder how to construct them and what
their minimum size is. 
The \Sparktope~project was motivated by these
questions. The LPs it produces are surely not minimal. However, for example,
is it possible to
find an LP to solve the maximum matching problem considered here which has
$o(n^5)$ constraints?
}

Although we have concentrated on converting algorithms for problems in $\Pp$
into polynomially sized LPs the method and software described here will work
for any algorithm that can be expressed in \Sparks. For example, consider the
travelling salesman problem. The convex hull of all TSP tours in the 
complete graph $K_n$ is called the TSP polytope. It is known that this polytope
has more than $n!$ facet defining inequalities (see, e.g., \cite{ABCC07}). 
However a dynamic programming
algorithm due to Held and Karp \cite{HK62} runs in $O(2^n n^2)$ time and
$O(2^n n)$ space. So the LP formulation produced by \Sparktope~based on
this algorithm has size $O(4^n n^3 \log n)$ which
is asymptotically smaller than the one based on the TSP polytope.
\change{
Is there an asymptotically smaller sized LP that can solve the travelling salesman problem?
}

\section*{Acknowledgements}
We would like to thank Bill Cook and Hans Tiwary for helpful discussions.
In particular the former suggested we consider the Held-Karp algorithm
and the latter the makespan problem.
Two referees provided valuable comments which helped us improve the manuscript.
This research is supported by 
the JSPS and NSERC.

\bibliographystyle{tfs}
\bibliography{sp}

\begin{appendices}

\section{\texttt{mm.spk} : sparks code}
\label{spkcode}

\includegraphics[width=\textwidth]{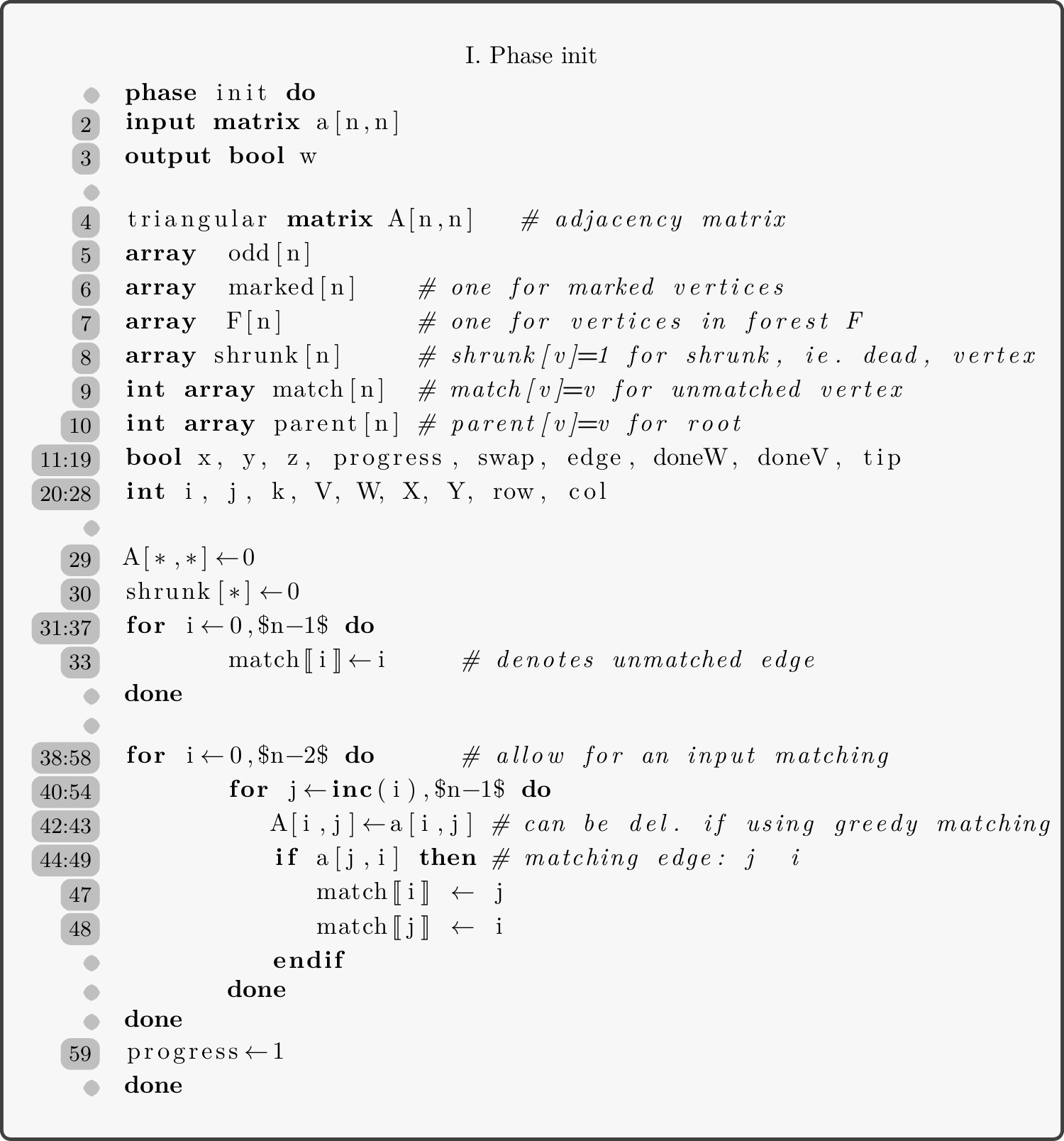}  

\includegraphics[height=\textheight]{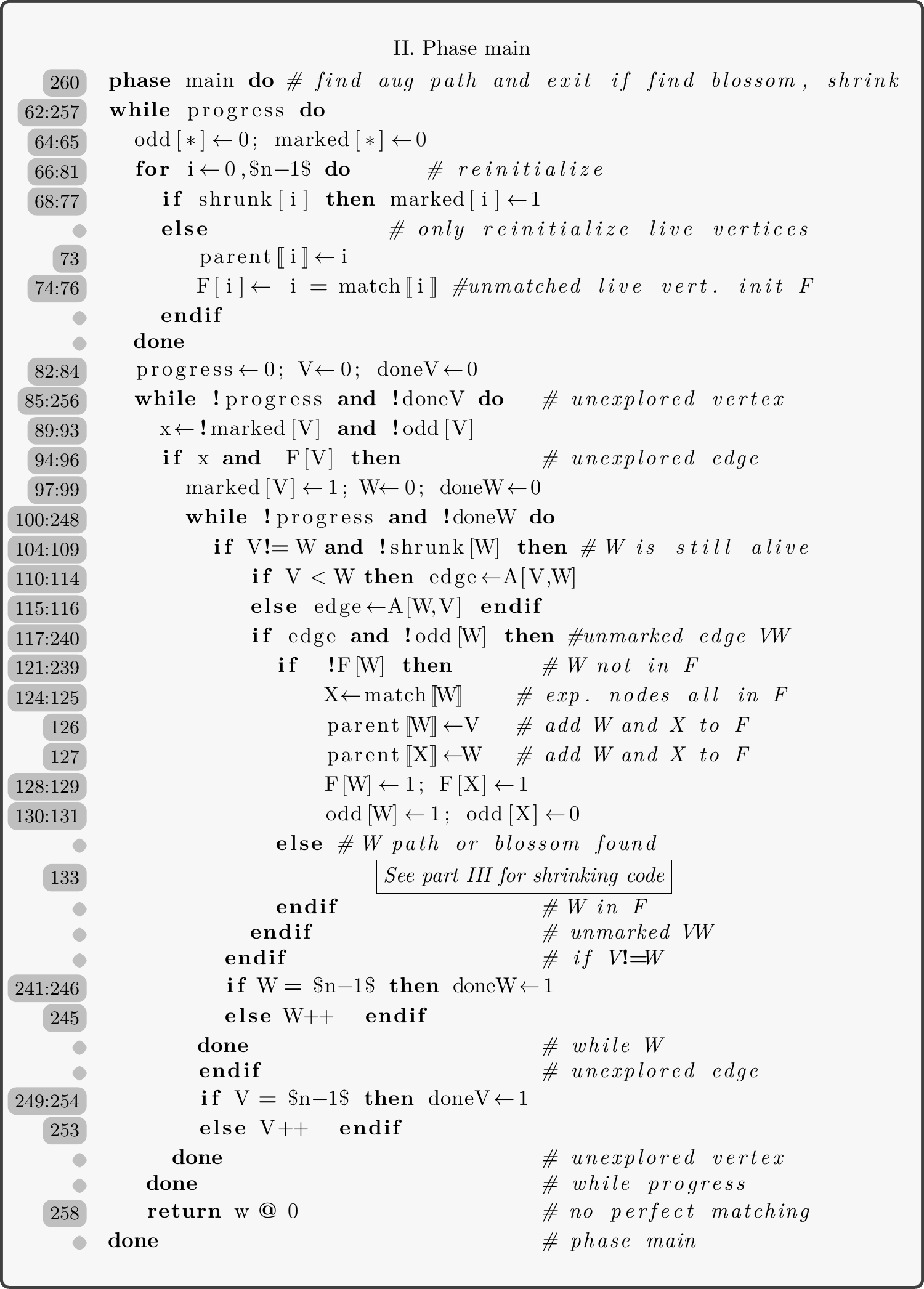}

\includegraphics[height=\textheight]{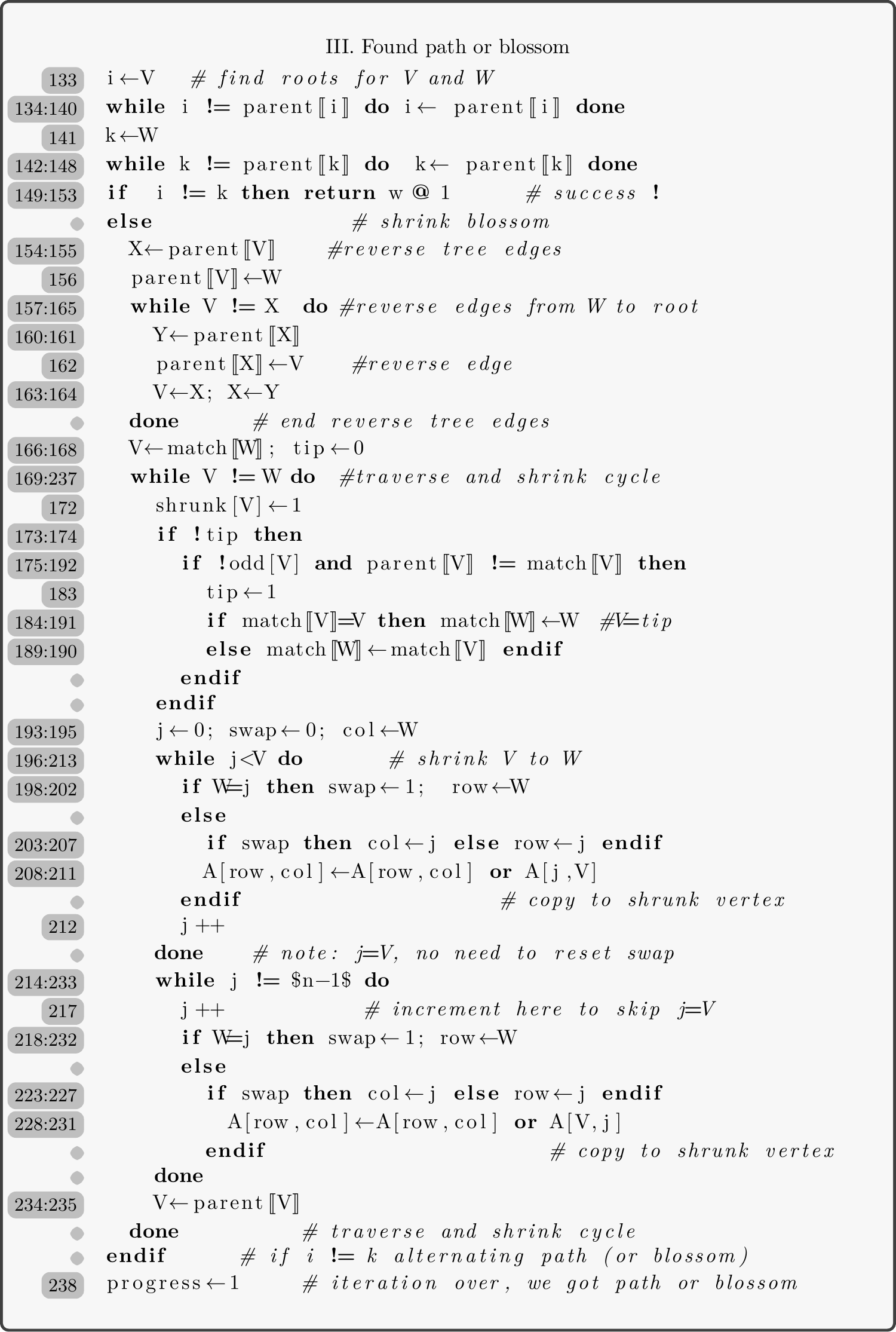}

\section{Sample inputs for maximum matching}
\label{inputs}

\begin{sparksbox}{}
wt8.in:
# a is an n by n binary matrix
# the upper triangle contains the adj matrix of a graph
# the lower triangle contains a (partial) matching 10, 32, 54 

matrix a[8,8] <- {{0,1,1,1,0,1,1,1},
                 {1,0,1,0,0,0,0,0},
                 {0,0,0,1,1,1,1,1},
                 {0,0,1,0,0,0,0,0},
                 {0,0,0,0,0,1,1,0},
                 {0,0,0,0,1,0,1,0},
                 {0,0,0,0,0,0,0,0},
                 {0,0,0,0,0,0,0,0}}
\end{sparksbox}

\begin{sparksbox}{}
wt8a.in:

matrix a[8,8] <- {{0,1,1,1,0,1,1,1},
                 {1,0,1,0,0,0,0,0},
                 {0,0,0,1,1,1,1,1},
                 {0,0,1,0,0,0,0,0},
                 {0,0,0,0,0,1,1,1},
                 {0,0,0,0,1,0,1,0},
                 {0,0,0,0,0,0,0,0},
                 {0,0,0,0,0,0,0,0}}
\end{sparksbox}

\end{appendices}
\end{document}